# Measurement theory of a density profile of small colloids around a large colloid: Superposition of the radial distribution functions

**Ken-ichi Amano and Kota Hashimoto**

*Department of Energy and Hydrocarbon Chemistry, Graduate School of Engineering, Kyoto University, Kyoto 615-8510, Japan.*

Author to whom correspondence should be addressed: Ken-ichi Amano.

Electric mail: amano.kenichi.8s@kyoto-u.ac.jp

**ABSTRACT**

We propose a transform theory for calculating a density profile of small colloids around a large colloid from a force curve between the two-large colloids. In the colloid solution, there are many small colloids and two or several large colloids. The force curve between the two-large colloids can be measured by laser tweezers. In this letter, the transform theory is derived in detail, where a superposition approximation of the radial distributions of the density profiles and rigid-body approximation are introduced. In our opinion, if the experimental condition is satisfied, the transform theory can be used not only for the laser tweezers, but also for surface force apparatus and colloid probe atomic force microscopy. Furthermore, the transform theory is to calculate a density profile of micelles around a large spherical surface.

**MAIN TEXT**

Recently, the laser tweezer is an important tool for detecting subtle force between two-small substances. For example, stepwise movements of a kinesin [1] and a myosin [2,3] are measured by the laser tweezer. Moreover, forces occurring while stretching (i.e., mechanical unfolding) a DNA chain [4] and a protein [5,6] are also measured by the laser tweezer. The laser tweezer has revealed many mechanisms of these dynamics and properties. Also in a colloid solution, the laser tweezers are used, where the potential of mean force (it readily be transformed into the mean force) between two-large colloids has been measured [7]. The colloid solution includes the two- or several-large colloids and many-small colloids in the aqueous solution. The force curve between the two-large colloids is oscillatory and the layer spacing is nearly equal to the diameter of the small colloid. In addition, due to the salt concentration in the experiment is at a proper quantity, the interactions between colloids are approximated by rigid-like and the aqueous solution is inert background. As a similar experiment, mean force between two cylindrical solids has been measured by using surface force apparatus (SFA), where micelles are immersed in the solvent [8].

In this letter, we propose a theory that transforms the force curve between two-large colloids into the density distribution of small colloids around the large colloid. The fundamental idea is based on our recent idea for SFA [9,10]. (In Ref. [9,10], the two probes are modeled as solids with flat surfaces.) Before the derivation of the transform theory, we illustrate the system considered here in Fig. 1. The large colloids 1 and 2 with radii $r_B$ are immersed in the solvent. They are caught by the laser tweezers [7]. In the solvent, there are many small colloids with radii $r_S$, number density of which is $\rho_0$. These colloids are modeled as hard spheres. The solvent is hypothecated as an inert background. The separation between the centers of the large colloids is expressed as $s$, the length between the center of the large colloid 1 and the center of the small colloid in contact with the large colloid 2 is expressed as $l$. $\theta$ assigns the orientation from the vertically upward line of the small colloid in contact with the large colloid 2.

The large colloid has a radial density distribution of the small colloids around itself. When the large colloids 1 and 2 approach, the radial density distributions overlap. The overlap makes the complex distribution of the small spheres between the large colloids 1 and 2. However, in a simple form, the overlap can be modeled by Kirkwood superposition approximation [11-13]:

$$\rho_0 g_W = \rho_0 g_1 g_2, \tag{1}$$

where $g_W$, $g_1$, and $g_2$ represent the normalized density distributions of the small colloids within the whole system, around the large colloids 1 and 2, respectively ($g_1$ and $g_2$ represent that in the bulk system). In a theory of the rigid wall, the mean force acting on the large colloid 2 depends only on the contact density of the small colloids around itself [9,14]. Therefore, the mean force $f$ (*more specifically, it may be called colloid-induced mean force*) is expressed as

$$f(s) = 2\rho_0 k_B T \pi r^2 \int_0^{\pi/2} g_1(l(\theta,s)) g_{2C} \sin\theta \cos\theta d\theta - f_B, \qquad (2)$$

where $r \equiv r_B + r_S$ and $g_{2C}$ represents the value of $g_2$ at the contact. $f_B$ (constant) is the mean force acting on the backside (right side) of the large colloid 2. It is expressed as

$$f_B = 2\rho_0 k_B T \pi r^2 \int_0^{\pi/2} g_{2C} \sin\theta \cos\theta d\theta = \rho_0 k_B T g_{2C} \pi r^2. \qquad (3)$$

Here, $l$ can be written as

$$l = \sqrt{r^2 + s^2 - 2sr \sin\theta}. \qquad (4)$$

The differentiation of $l$ with respect to $\theta$ is written in the form:

$$\frac{dl}{d\theta} = -\frac{\sqrt{4s^2 r^2 - (s^2 + r^2 - l^2)^2}}{2l}. \qquad (5)$$

Moreover, $\sin\theta$ and $\cos\theta$ are written as

$$\sin\theta = \frac{s^2 + r^2 - l^2}{2sr}, \qquad (6)$$

$$\cos\theta = \sqrt{\frac{4s^2 r^2 - (s^2 + r^2 - l^2)^2}{4s^2 r^2}}. \qquad (7)$$

Thus, the mean force is rewritten as

$$f(s) = \rho_0 k_B T \pi g_{2C} \int_{s-r}^{\sqrt{s^2+r^2}} g_1(l) \frac{l(s^2 + r^2 - l^2)}{s^2} dl - f_B, \qquad (8)$$

or

$$\frac{f(s) + f_{\text{B}}}{\rho_0 k_{\text{B}} T \pi g_{2\text{C}}} = \int_{s-r}^{\sqrt{s^2+r^2}} \frac{l(s^2 + r^2 - l^2)}{s^2} g_1(l) dl. \tag{9}$$

It can be seen that Eq. (9) is in the form of a matrix calculation as follows:

$$\boldsymbol{F}^* = \boldsymbol{H}\boldsymbol{G}_1, \tag{10}$$

where $\boldsymbol{F}^*$ corresponds to left-hand side of Eq. (9), whose (main) variable is *s*. $\boldsymbol{G}_1$ and $\boldsymbol{H}$ correspond to $g_1(l)$ and the other parts, respectively. $\boldsymbol{H}$ is a square matrix whose variables are *l* and *s*, however, its lower right area is composed of a square unit matrix. Subsequently, $\boldsymbol{G}_1$ is numerically calculated by substituting an arbitrary-initial value of $g_{2\text{C}}$ and by using, for example, the inverse matrix of $\boldsymbol{H}$. When the output of $g_{1\text{C}}$ (the contact value of $g_1$) is equal to $g_{2\text{C}}$, that is when $\Delta g = g_{1\text{C}} - g_{2\text{C}} = 0$, the numerical calculation is finished. The convergence of $\Delta g$ is operated by, for example, a bisection method.

  As a supplement, we would like to mention that the three-dimensional (3D) density distribution of the small colloids between the two-large colloids can be calculated by substituting $g_1$ and $g_2$ into Kirkwood superposition approximation (Eq. (1)). Thus, the transform theory is possible to calculate the two things: (I) radial-density distribution of the small colloids around the large colloid; (II) 3D-density distribution of the small colloids between the two-large colloids.

  We verified the transform theory derived here by using a computer. At first, the density distribution of the small colloids around the large colloid is calculated by using one-dimensional Ornstein-Zernike equation coupled with hypernetted-chain closure (1D-OZ-HNC), previously. Next, the input datum (force curve between the two large colloids) is calculated also by 1D-OZ-HNC. (The system is modeled as rigid and the solvent here is inert background.) The input datum is substituted into the transform theory, and then the density distribution of the small colloids around the large colloid (i.e., output) is obtained. We have found that the output is similar to the previously calculated density distribution (not shown). Thus, the validity of the transform theory has been confirmed.

  In summary, we have derived the transform theory that transforms the force curve between the two-large colloids into the density distribution of the small colloids around the large colloid. In addition, we proposed the simple method for estimating the 3D density distribution of the small colloids between the two-large colloids. We have briefly verified the transform theory, and concluded that the theory is valid. In the near future, we will conduct detailed verification of the theory and find applicable range of it.

**ACKNOWLEDGEMENTS**

We appreciate Tetsuo Sakka (Kyoto University), Naoya Nishi (Kyoto University), and Masahiro Kinoshita (Kyoto University) for the useful advice, discussions, and data. This work was supported by "Grant-in-Aid for Young Scientists (B) from Japan Society for the Promotion of Science (15K21100)".

# FIGURE

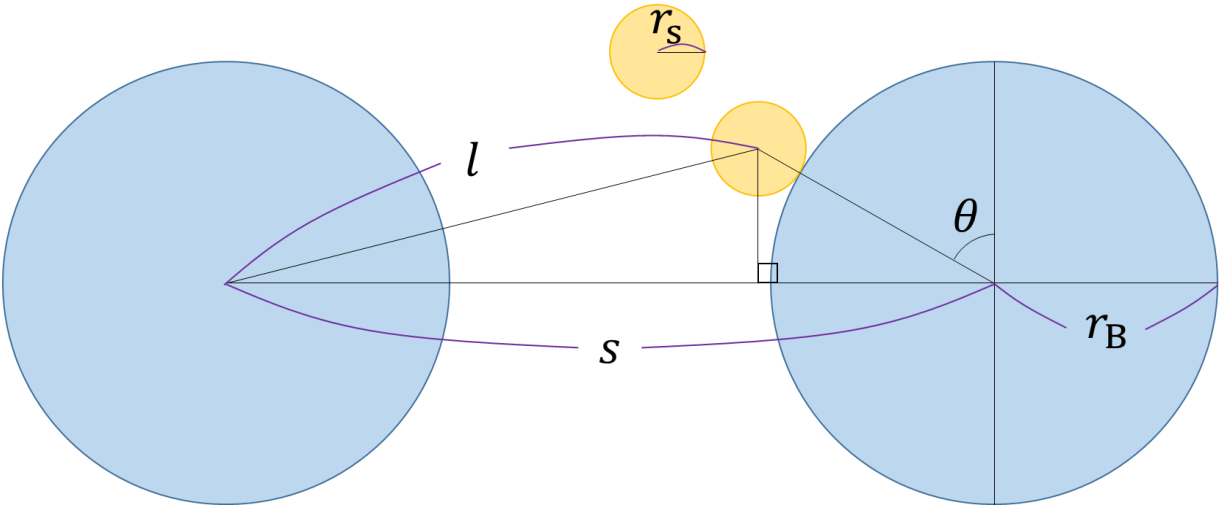

Fig. 1: The system configuration.